\documentclass[conference]{IEEEtran}
\IEEEoverridecommandlockouts
\usepackage{cite}
\usepackage{amsmath,amssymb,amsfonts}
\usepackage{algorithmic}
\usepackage{graphicx}
\usepackage{textcomp}
\usepackage{xcolor}
\usepackage{array}
\usepackage{multirow} 
\usepackage{hyperref}

\begin{document}

\title{Quantization for OpenAI's Whisper Models: A Comparative Analysis\\}

\author{\IEEEauthorblockN{1\textsuperscript{st} Allison Andreyev}
\IEEEauthorblockA{\textit{Independent Researcher} \\
Washington DC, United States \\
allisonmandreyev@gmail.com}
}

\maketitle
\begin{abstract}
Automated speech recognition (ASR) models have gained prominence for applications such as captioning, speech translation, and live transcription. This paper studies Whisper and two model variants: one optimized for live speech streaming and another for offline transcription. Notably, these models have been found to generate hallucinated content, reducing transcription reliability. Furthermore, larger model variants exhibit increased latency and pose challenges for deployment on resource-constrained devices. This study analyzes the similarities and differences between three Whisper models, qualitatively examining their distinct capabilities. Next, this study quantifies the impact of model quantization on latency and evaluates its viability for edge deployment. Using the open source LibriSpeech dataset, this paper evaluates the word error rate (WER) along with latency analysis of whispercpp using 3 quantization methods (INT4, INT5, INT8). Results show that quantization reduces latency by 19\% and model size by 45\%, while preserving transcription accuracy. These findings provide insights into the optimal use cases of different Whisper models and edge device deployment possibilities.
All code, datasets, and implementation details are available in Appendix Sec. \ref{sec: appendix-github}.
\end{abstract}

\section{Introduction}
\label{sec: int}
The rise of large language models (LLMs) has enabled advancements across multiple communication modalities, including speech processing. \texttt{Whisper} is an automated speech recognition (ASR) system developed by OpenAI, designed for applications such as speech transcription, live translation, and captioning \cite{doi:10.1177/03400352241310534}.
The model has been trained with 680,000 hours of audio data, far surpassing the magnitude of training data used for ASR models. Furthermore, as training data has been divvied into 97 total languages, \texttt{Whisper} is compatible to act as a translation machine, frequently found to perform better than LLM-based ASR models \cite{song2024comparativestudyllmbasedasr}. This robust approach has made \texttt{Whisper} one of the leading ASR models in both research and practical applications, and allows \texttt{Whisper} to be amongst the top performers across many applications of speech processing, such as through translation, transcription, speech recognition, and zero-shot evaluation \cite{radford2022robust}. Additionally, due to \texttt{Whisper}'s open support for quantization e.g \texttt{whispercpp} --a lightweight model with built in quantization features--, it led me to use \texttt{Whisper} as the focus of this study \cite{whispercpp}. Currently, \texttt{Whisper} contains five versions: tiny, small, base, medium, and large. Due to its accuracy and easy of use, \texttt{Whisper} is becoming increasingly popular in both research and commercial applications \cite{liu2024whisper}. However, these models are not always accurate when transcribing speech, and some transcriptions have been found to contain hallucinations. Moreover, larger model variants exhibit increased latency and computational demands, making deployment on resource-constrained devices more challenging. \cite{10.1145/3630106.3658996} find that roughly 1\% of audio transcriptions by \texttt{Whisper} contained entire hallucinated phrases or sentences, while also finding 38\% of hallucinations included harms such as violence, inaccuracies or false authority. While prior research has explored fine-tuning strategies to enhance \texttt{Whisper}’s performance \cite{liu2024whisper, 10.1121/10.0024876, johansson2024automatic, 10447492,rijal2024whisperfinetuningnepalilanguage, timmel2024finetuningwhisperlowresourcelanguages, 10581773}, the impact of quantization on model size reduction and latency optimization remains underexplored. 

\subsection{Contributions and Organization}

This paper addresses this research gap with a few contributions:

\begin{itemize}
    \item Evaluating the capabilities of \texttt{Whisper} and 2 variants: \texttt{Whisper\_Streaming} \& \texttt{whisper-timestamped}, with an emphasis on their similarities and differences.
    \item Establishing and defining quantization techniques and relevant hardware considerations applicable to \texttt{Whisper} models.
    \item A qualitative review on usage between the \texttt{Whisper} and \texttt{Whisper\_Streaming} \& \texttt{whisper-timestamped}
    \item Examining model performance in terms of word error rate, processing speed, and latency, relative to model size, version, runtime, and quantization approach.
    \item Summarizing both qualitative and quantitative results from two experimental evaluations.
\end{itemize}

I describe current research results and limitations in Sec. \ref{sec: litreview}. Sec. \ref{sec: modelComp} qualitatively compares Whisper and its two variants. In Sec. \ref{sec: exp1}, I conduct a model performance experiment, followed by a discussion of qualitative results in Sec. \ref{sec: results1}. Sec. \ref{sec: exp2} compares the Whisper base model size with its performance on the LibriSpeech audio dataset. Sec. \ref{sec: QuantIntro} introduces quantization and its applications for Whisper, while Sec. \ref{sec: QuantComp} compares the performance of whispercpp based on run type and quantization method. Sec. \ref{sec: exp3} evaluates the impact of three quantization methods on whispercpp's accuracy, latency, and model size. Finally, Sec. \ref{sec:apps} discusses the practical applications of this research, with a conclusion in Sec. \ref{sec:conc}.

\subsection{Note on Naming Conventions}

\label{sec: Note}
In this work, I explore several variants of the \texttt{Whisper model}. The naming conventions across these variants, developed by external organizations or individuals, are inconsistent, with some adopting different conventions:

\begin{itemize}
    \item \textbf{Whisper}: Standard model developed by OpenAI
    \item \textbf{whispercpp}: C++ implementation of the Whisper model
    \item \textbf{whisper-timestamped}: Version with added timestamping functionality
    \item \textbf{Whisper\_Streaming}: Version adapted for streaming
\end{itemize}

For consistency in this document, I will refer to these variants using the most common or descriptive form of the name. Please note that the developers may use different naming conventions across various implementations.
\section{Literature Review}
\label{sec: litreview}

So far, several studies have been conducted on quantizing LLMs and ASR speech transcription.
\cite{radford2022robust} evaluate the \texttt{Whisper} model, describing it as an encoder-decoder transformer model, in which a decoder predicts a corresponding text caption for each audio segment. They also note that \texttt{Whisper} was trained on a large and diverse dataset, which explains why its performance on a specific dataset may not be as high as a model trained on only one kind \cite{radford2022robust}. \cite{song2024comparativestudyllmbasedasr} evaluate \texttt{Whisper} model capabilities, and contrast it with LLM-based ASR models. \cite{song2024comparativestudyllmbasedasr} explain how LLM-based ASR models use a speech encoder to process speech and generate embeddings which are passed into a decoder-only LLM. In their experimentation, they find that the performance of LLM-based ASR models correlates positively with the proficiency of the LLM in the language being recognized, posing a limitation for LLM based ASR models. 
Additionally, \cite{barański2025investigationwhisperasrhallucinations} studies \texttt{Whisper} hallucinations in audio transcriptions, noting that \texttt{Whisper} has a tendency to generate and produce incorrect repetitions of recognized text. Hallucinations refer to the generation of transcriptions that include fabricated or incorrect information that was not part of the original speech. Their experiment finds several offensive hallucinations produced in data transcription and notes that audio length significantly affects the error rate and audio content had minimal relation on hallucination. \cite{barański2025investigationwhisperasrhallucinations}. Hallucinations pose a challenge for \texttt{Whisper}, however this is a issue that could be addressed with model quantization \cite{zhao2024speakeradaptationquantisedendtoend}, a method which has been previously found to decrease the WER, improving model accuracy \cite{zhao2024speakeradaptationquantisedendtoend}.
\cite{gholami2021surveyquantizationmethodsefficient} evaluate current quantization strategies, noting that moving from FP to INT quantization holds potential to reduce latency and memory footprint by a factor of 16x. Several studies had conducted experiments quantizing ASR models. For example, \cite{zhen2022sub8bitquantizationondevicespeech} evaluates quantization methods effects on model latency, finding that quantization creates faster model inference and allows for model deployment on portable devices, on which a stable network connection could be limited. However, the study mainly evaluates INT8 quantization, which is only one kind of integer quantization and may not speak for all methods such as INT4 or INT5, respectively. 
Meanwhile, \cite{9747552} discusses quantizing ASR models: they find that neural network architectures such as \texttt{Whisper} perform poorly on edge hardware due to computation requirements, and note that prior research on quantizing ASR models is limited. Notably, QAT requiring training and validation data during quantization may not always be available due to privacy or security issues, forming a limitation for quantization models which require QAT. 
\cite{zhao2024speakeradaptationquantisedendtoend} evaluate quantization methods on \texttt{Whisper} accuracy and model size, relying on the P4Q quantization strategy, which utilizes block-wise N4 quantization applied to the model's primary weights. The study evaluates their methods on 4 quantized \texttt{Whisper} models, demonstrating improvement in latency and accuracy, with a 15.1\% WER reduction for quantized \texttt{Whisper}. However, similar to \cite{zhen2022sub8bitquantizationondevicespeech}, this study doesn't evaluate more than 1 type of quantization, which limits the experiment in terms of deducing a pattern or relationship between quantization methods and model performance. Additionally, \cite{zhen2022sub8bitquantizationondevicespeech} proposes a general quantizer which uses a quantization scheme with floating point (FP) and backward-pass quantization aware training (BP-QAT). To evaluate this model, they use \texttt{Whisper} and the LibriSpeech dataset and benchmark accuracy using a standard WER, finding it improves by up to 5.7\% with their quantization methods. This method, however, is only evaluated on one model and version of \texttt{Whisper}, so its accuracy and results are not confirmed for other sizes. 
Research on quantization has found it could bear benefits to model accuracy, latency, and deployment opportunities \cite{gholami2021surveyquantizationmethodsefficient, 9747552, zhao2024speakeradaptationquantisedendtoend, zhen2022sub8bitquantizationondevicespeech}. Quantizing the model could benefit users who don't have stable internet access, or need to use the model on a mobile device. The translation and transcription features of the model pose as key resources needed by the hard of hearing community along with language barriers. However, current research is limited in terms of quantization strategies applied, and latency categories studied. Furthermore, up until now, few studies have compared \texttt{Whisper} and its variants  -\texttt{whisper-timestamped} and \texttt{Whisper\_Streaming}-- and the implications of each model individually. By analyzing \texttt{Whisper}'s variants, greater insights can be analyzed about the model backend and differentiating factors. For example,  \texttt{Whisper\_Streaming} uses self-adaptive latency \cite{machacek-etal-2023-turning}, possibly affecting how latency is impacted by quantization. Meanwhile,  \texttt{whisper-timestamped} uses a Dynamic Time Warping (DTW) approach (unique to  \texttt{whisper-timestamped}) which allows for improved timestamp accuracy, while also featuring confidence scores for each word. Furthermore, \texttt{whisper-timestamped} is able to process longer files with little additional memory usage compared to the \texttt{Whisper} Base model \cite{lintoai2023whispertimestamped, JSSv031i07}. This difference in memory handing and individual word confidence scores pose interesting new research directions about this model. In all, the \texttt{Whisper} model comparisons can clearly conclude on variant applications and limitations, which play a key role in further research and development. 

This paper aims to address these limitations by providing a comprehensive evaluation of the 3 \texttt{Whisper} variations while running a quantization experiment using 3 methods of integer quantization with a comprehensive model accuracy and latency analysis.\\

\section{Model Comparative Analysis}
\label{sec: modelComp}

\subsection{Whisper}

Whisper, developed by OpenAI, is an ASR model capable of speech transcription, translation, and language identification. When transcribing, the model detects voice activity and attenuates background noise or music. When setting up \texttt{Whisper}, its dependencies 
\textit{PyTorch \& ffmpeg command line tool} must also be installed. OpenAI has recently made an API version that can be imported into any Python file as a module for personal modification. However, it is noted \texttt{Whisper} is not designed for real-time transcription, and is only made to process audio with at least 1 full sentence, which is preferably less than 30 seconds in length.  

\subsection{Whisper Timestamped}

\texttt{whisper-timestamped} is a version of \texttt{Whisper} that creates word timestamps and more exact estimations on speech segments using a Dynamic Time Warping approach \cite{lintoai2023whispertimestamped}. That way, start and end time estimations for speech are more accurate and each word is processed individually, receiving its own confidence score. By employing this approach, \texttt{whisper-timestamped} can process longer files with minimal additional memory overhead. OpenAI offers a 9GB docker file and light installation of CPU, and \texttt{whisper-timestamped} can also be used as a Python module. \texttt{whisper-timestamped} has several output formats:

\begin{itemize}
    \item Outputs data in JSON format with: detailed timestamp data, language detection, confidence score
    \item CSV, SRT, VTT, TSV file
    \item Into specified output directory, 'verbose' mode
\end{itemize}

Additionally, computation and confidence scores can be enabled and disabled for each word, and the user can choose whether punctuation should be committed \cite{Giorgino2009-os, machacek-etal-2023-turning, radford2022robust}.

\subsection{Whisper Streaming}

\texttt{Whisper\_Streaming} is an optimized variant of \texttt{Whisper} designed for real-time speech transcription and translation. Typically, \texttt{Whisper\_Streaming} can transcribe live speech with a 3.3 second latency. On top of dependencies required for \texttt{Whisper}, \texttt{Whisper\_Streaming} requires the Libra Sound File, a sound processing library, and requires installing the \texttt{Whisper} back end and the OpenAI API. \texttt{Whisper\_Streaming} comes with 4 simulation modes:

\begin{itemize}
    \item \textbf{Start\_at}: Starts processing audio at a time provided by user.
    \item \textbf{Offline}: Processes the full audio file once in offline mode, and then finds the lowest word error ratio.
    \item \textbf{Comp\_unaware}: Timer that measures processes/events; does not count compute time, meant to lower latency bounds and get 'true' latency.
    \item \textbf{Default usage}
\end{itemize}

Next, there were a few key similarities and differences between \texttt{Whisper\_Streaming} and the other 2 models.

Text and debug variables are outputted as soon as that piece of the speech is processed, doing the transcription live rather than outputting final data at the end. The model takes a significantly longer time to process longer audio; however, similar to \texttt{whisper-timestamped}, \texttt{Whisper\_Streaming} offers several customization features in Python files and the command line, such as an offline mode, customizing buffer timing, when streaming starts, the language used, model, and minimum segment size (what size the buffer transcribes at a time). Longer audio files need to be split into small pieces and then merged. In low latency streaming mode, words can be split in the middle. Unlike the other models, \texttt{Whisper\_Streaming} does not do sentence segmentation: it instead makes word-level timestamps. The model processes the new audio segment twice before finalizing, updates the buffer to the timestamp with confirmed audio segment. Limits processing buffer window \& reprocesses the confirmed sentence time stamps before moving on to the next speech piece, this is because the objective is to limit buffer size and increase efficiency for longer audio segments, while still ensuring accuracy.

\subsubsection{Limitations}

Due to lots of terminal output, it was difficult to see the full text transcript. Noted, this is mitigated however, by the model storing the transcript in a separate txt file. Based on this output, \texttt{Whisper\_Streaming} is not ideal for pre-recorded audios due to its buffering method and accumulation of text data, making the terminal harder to sort. Additionally, for increased accuracy, each audio segment has one word processed at a time, which can cause lag in the software.\\\\
All can be used as a Python module using its API.

\section{Model Performance Experiment}
\label{sec: exp1}

To evaluate the accuracy and latency of the base \texttt{Whisper} models, this study utilizes 25 audio files from the LibriSpeech dataset, comprising both clean and challenging speech samples \cite{panayotov2015librispeech}.
All models were executed in a standardized virtual environment using Jupyter Notebook on an HP Envy CPU to ensure consistency. From this experiment, qualitative results were derived on the performance of all 3 models relative to each other.
\section{Qualitative Results}
\label{sec: results1}

\subsection{Whisper Timestamped}

\texttt{whisper-timestamped} provided additional personalization compared to \texttt{Whisper}, such as specifying a file output directory. A function in the Python module of  \texttt{whisper-timestamped} also takes several parameters that allow for further customization on the way the speech is transcribed. This model breaks off speech into segments, and then segments into words, sharing timestamps and confidence scores for each individually. While \texttt{Whisper} provided timestamps for each sentence, \texttt{whisper-timestamped} provided a timestamp for each word and confidence score for each sentence, phrase, and word, separately.
Confidence is rated on a scale from 0.00 to 1.0. Notably, the confidence score and processing speed stayed the same for the longer and more complicated pieces of text.
\texttt{whisper-timestamped} also has a few unique traits:

\begin{itemize}
    \item Features a progress bar at the top of the output, with FPS and percentage processed.
    \item Output by default is in JSON format.
    \item Specifying a language when translating would remove the output featuring language probability.
\end{itemize} 

Between model sizes, the largest model processes text at about 400 FPS and the smallest model processes almost 2000 FPS. Confidence levels increased dramatically for each word between the tiny and large models. On average, the base model took about 10 seconds to transcribe speech. All models did not interpret intonation to structure sentences and capitalization properly.

\subsection{Whisper}

Using \texttt{Whisper} without the timestamped feature provided me with some timestamps with text that were particularly long. These timestamps were typically a range of a few seconds (for each sentence). By default, the model provided 5 output files: JSON, vtt, srt, txt, tsv
All models would use commas, punctuation, and capitalization correctly, they were able to apply grammatical rules to a sentence, which differentiated it from other speech transcription models. 
For example, a sentence such as “Bob’s dogs were happily playing with Cat” would’ve translated properly with \texttt{Whisper} [capitalizing ‘C’ in ‘cat’ to make a name], but may have translated to “Bobs dogs were happily playing with cat” on non-AI powered transcription devices.
The time it took for the models to compute the language and text was the same for the more complex speech as the more clean speech. (About 10 seconds for the base model) Unlike \texttt{whisper-timestamped}, \texttt{Whisper} did not output: confidence scores, timestamps for each word, and language probability.
For both models, performance on the test and development set performance was similar. Throughout usage, there were some notable semantics utilized by the model, such as intonation to determine capitalization and sentence structure. The model goes through 2 layers of transcription, and it adjusts transcription as it goes, predictions of text may change in a buffer as they’re being double-checked or the next part is being listened to.

\section{Model Size vs. Performance Experiment}
\label{sec: exp2}

Using 10 distinct recordings from the ‘test-clear’ and ‘test-other’ data sets from Librispeech, each audio segment was manually timestamped and compared with the timestamps provided by \texttt{Whisper\_Streaming} and \texttt{whisper-timestamped}'s base versions. Timestamps went up to the centiseconds (cs).

\subsection{Whisper Streaming}

\texttt{Whisper\_Streaming} demonstrated strong accuracy, with automatically generated timestamps deviating no more than 0.5 seconds from manually recorded ones. However, the software would frequently start each timestamp from 0.00s, even though the words started being spoken at a later point in the audio recording. Additionally, it was a frequent pattern to notice the software undercount seconds needed to pronounce a phrase, being about 0.2s ahead of human-recorded timestamps most of the time.

\subsection{Whisper Timestamped}

\texttt{whisper-timestamped} had more distinct timestamps, with each word getting its separate time frame. It seemed to have the same level of accuracy as \texttt{Whisper\_Streaming} when it came to comparing the generated timestamps to human-recorded ones. The benefit of \texttt{whisper-timestamped} over \texttt{Whisper\_Streaming} was the increased precision of the timestamps and more detail. \texttt{whisper-timestamped} also started off with more accurate timestamps than \texttt{Whisper\_Streaming}, which usually started each segment at 0.00s.

\subsection{Observations}
\hbadness=10000  

The following table displays qualiative usage observations with \texttt{Whisper} based on model size.

\begin{table}[h]
\centering
\raggedright
\caption{Qualitative Whisper Usage Experience on LibriSpeech Datasets Based on Model Size and Speech Difficulty} 
\begin{tabular}{|l|p{3cm}|p{3cm}|}
\hline
\textbf{Model Size} & \textbf{Clean Speech} & \textbf{Challenging Speech} \\
\hline
\textbf{Tiny} & 
Quick output (< 10s), low GPU/CPU usage, inaccuracies with larger text or names, capitalization issues & 
Misses small background noises, e.g., “They worshiped” only “worship” heard \\
\hline
\textbf{Small} & 
10-20s output, best capitalization, good timestamp details & 
Similar to Medium, but 2x faster \\
\hline
\textbf{Medium} & 
20-40s output, similar accuracy to large model & 
N/A \\
\hline
\textbf{Large} & 
Long download (2GB), slow processing (up to a couple of minutes), punctuation and capitalization issues & 
Modifies structure to be grammatically correct while matching audio more closely \\
\hline
\end{tabular}
\label{tab:model_comparison_swapped}
\end{table}

\subsection{Results}

\texttt{Whisper\_Streaming} demonstrated strong accuracy, with automatically generated timestamps deviating no more than 0.5 seconds from manually recorded ones. The algorithm for both models was programmed to detect and check for trailing sounds like ‘s’ or ‘n’ correctly and performed well against human-labeled timestamps. \texttt{whisper-timestamped} prioritized precision and granularity, albeit at the cost of increased processing time compared to other models.
\section{Quantization with Whisper}
\label{sec: QuantIntro}

Quantization involves converting a neural network (NN) or large language model (LLM) to a lower-precision format, reducing memory requirements for deployment on resource-limited devices. Quantization maps continuous input values to discrete levels at the output. \texttt{Whisper}, being an audio LLM, has the ability to be quantized to be deployed and used on smaller devices with less computing power. Thus, the following experiment evaluates whether model accuracy and latency are affected positively through quantization techniques. 

\subsection{Hardware Support for Quantization}

The deployment of quantized models benefits from hardware accelerators that optimize the efficiency of quantization tasks. Key hardware platforms include:
\begin{itemize}
    \item \textbf{AMD and ARM CPUs:} Support for mixed-precision operations and 8-bit integer quantization, with notable examples including AMD Zen 4 and ARM Neoverse V1/V2 architectures.
    \item \textbf{Apple Silicon and NVIDIA GPUs:} Apple's chips (A17 Pro, M4) and NVIDIA's H100 GPU offer enhanced support for 8-bit integer quantization and tensor core optimization.
    \item \textbf{Intel CPUs and Qualcomm GPUs:} Intel Xeon processors feature support for 8-bit integer quantization via AMX, while Qualcomm Adreno GPUs provide optimization for mixed-precision tasks.
\end{itemize}

\section{Quantized Whisper CPU vs. GPU Performance}
\label{sec: QuantComp}

Using several quantization methods (Q4, Q5, and Q8), the base \texttt{Whisper} model was quantized to compare different implementation methods, experiences, accuracy, and observed differences.

The following table describes the difference in total run time, divided into several components based on quantization and runtime type (CPU versus GPU).

\begin{table}[ht]
\raggedright
\caption{Whisper Baseline vs. Quantized Compute Times based on Hardware}
\begin{tabular}{|>{\raggedright\arraybackslash}p{1.2cm}|p{1.35cm}|p{1.35cm}|p{1.35cm}|p{1.35cm}|}
\hline
\multirow{2}{*}{\textbf{Time}} & \multicolumn{2}{c|}{\textbf{whispercpp (GPU)}} & \multicolumn{2}{c|}{\textbf{whispercpp (CPU)}} \\
\cline{2-5}
 & \textbf{Standard} & \textbf{Quantized} & \textbf{Standard} & \textbf{Quantized} \\
\hline
Load & 123.58 ms & 66.54 ms & 162.27 ms & 94.51 ms \\
Mel & 43.29 ms & 51.26 ms & 80.58 ms & 80.88 ms \\
Sample & 2.03 ms/run & 1.47 ms/run & 1.84 ms/run & 1.87 ms/run \\
Encode & 4604.79 ms/run & 5934.99 ms/run & 6468.15 ms/run & 8612.49 ms/run \\
Decode & 226.40 ms/run & 9.75 ms/run & 12.56 ms/run & 11.16 ms/run \\
Batchd & 9.94 ms/run & 7.76 ms/run & 7.94 ms/run & 9.10 ms/run \\
Prompt & 0.00 ms/run & 0.00 ms/run & 0.00 ms/run & 0.00 ms/run \\
Total & 6786.58 ms & 7414.24 ms & 8033.38 ms & 10380.28 ms \\
\hline
\end{tabular}
\end{table}

\section{Quantization for Optimization Experiment}
\label{sec: exp3}

Next, the experiment evaluates how accuracy and latency are affected by the quantization of the \texttt{Whisper} models. Reference the Appendix
\ref{sec:appendix_hwspecs} for hardware specifications.

For my audio data, I used the open source Librispeech ASR \cite{panayotov2015librispeech} dataset, using the first 10 audio files provided. These audio transcriptions then determine how quantization methods (Q4, Q5, Q8) affect model speed and accuracy. This study evaluates the WER and accuracy using \cite{ins8ai_2023}, a model which uses components of huggingface-evaluate and openai-whisper projects for WER calculation. The following table records key observations:

\begin{table}[h!]
\raggedright
\caption{WER, Model Size, and Latency based on Quantization Method}
\resizebox{8.8cm}{!}{
\begin{tabular}{|l|l|l|l|l|}
\hline
\textbf{Metric} & \textbf{Whisper CPP Base Model} & \textbf{INT5} & \textbf{INT4} & \textbf{INT8} \\ \hline
Word Error Rate & 0.0199 & 0.0199 & 0.0159 & 0.0199 \\ \hline
Accuracy & 98.0\% & 98.0\% & 98.4\% & 98.0\% \\ \hline
Model Size & 141.11MB & 52.75MB & 44.33MB & 77.99MB \\ \hline
Avg Latency & 10.64s & 11.11s & 10.55s & 9.02s \\ \hline
\end{tabular}
}
\end{table}
\section{Applications}
\label{sec:apps}

The results of this study could have profound implications for resource-constrained environments, such as mobile devices, IoT, and embedded systems. By reducing the memory footprint and maintaining high accuracy, quantized \texttt{Whisper} models could provide real-time transcription services, enable low-latency captioning, and improve accessibility for users with hearing impairments. The deployment of the Whisper models onto portable devices sets a high benchmark for transcription, translation, and speech detection services available for commercial use.
\section{Conclusion}
\label{sec:conc}

The study concludes that quantization is a viable method for reducing model size and improving deployment efficiency without sacrificing accuracy or latency. This experiment reduces model size by up to 45\% while maintaining the same WER and decreasing latency by 19\%. These results support the feasibility of \texttt{Whisper} on smaller devices, suggesting that \texttt{Whisper} can be effectively deployed on resource-limited devices such as smartphones and IoT systems, making real-time ASR more accessible and efficient. Extending this research to other ASR models could enhance the scalability of audio-based AI applications. Future work could explore additional quantization techniques, optimize hardware deployment strategies, and investigate trade-offs in real-time performance. As ASR models scale, optimizing the trade-off between accuracy, efficiency, and real-time performance will be critical for next-generation AI deployment.

\bibliographystyle{abbrv-doi}
\bibliography{refs}

\appendix

\subsection{}
\label{sec: appendix-github}
\href{https://github.com/allisonandreyev/WhisperQuantization}{https://github.com/allisonandreyev/WhisperQuantization}

\renewcommand{\thesection}{\Alph{section}}
\subsection{}
\label{sec:appendix_hwspecs}
\footnotesize
\noindent
Cpuinfo Version: 9.0.0\\
Brand Raw: Intel(R) Xeon(R) CPU @ 2.20GHz\\
Hz Advertised Friendly: 2.2000 GHz\\
Hz Actual Friendly: 2.2000 GHz\\
Hz Advertised: (2200000000, 0)\\
Hz Actual: (2199998000, 0)\\
Arch: X86\_64\\
Bits: 64\\
Count: 2\\
Arch String Raw: x86\_64\\
L1 Data Cache Size: 32768\\
L1 Instruction Cache Size: 32768\\
L2 Cache Size: 262144\\
L2 Cache Line Size: 256\\
L2 Cache Associativity: 6\\
L3 Cache Size: 57671680

\end{document}